\documentclass[AMA,Times1COL]{WileyNJDv5}

\articletype{Article Type}%

\received{Date Month Year}
\revised{Date Month Year}
\accepted{Date Month Year}
\journal{Journal}
\volume{00}
\copyyear{2023}
\startpage{1}

\begin{document}

\title{Analysis of Stepped-Wedge Randomised Cluster Trial using a generalized pairwise comparison approach : a simulation study}

\author[1]{Yohan Bard}

\author[2]{Emilie Presles}

\author[3]{Marc Buyse}

\author[4]{Silvy Laporte}

\author[5]{Paul Zufferey}

\author[6]{Frederikus A. Klok}

\author[7]{Olivier Sanchez}

\author[8,9]{Francis Couturaud}

\author[4]{Edouard Ollier}

\authormark{BARD \textsc{et al.}}
\titlemark{\textsc{Analysis of Stepped-Wedge Randomised Cluster Trial using a generalized pairwise comparison approach : a simulation study}}

\address[1]{\orgdiv{SAINBIOSE}, \orgname{INSERM U1059}, \orgaddress{\state{Saint-Etienne}, \country{France}}}

\address[2]{\orgdiv{Clinical Investigation Center CIC 1408}, \orgname{CHU Saint-Étienne}, \orgaddress{\state{Saint-Étienne}, \country{France}}}

\address[3]{\orgdiv{International Drug Development Institute (IDD)}, \orgname{Biostatistics, Louvain la Neuve and University of Hasselt, I-BioStat}, \orgaddress{\state{Hasselt}, \country{Belgium}}}

\address[4]{\orgdiv{Clinical Research Unit}, \orgname{
University Hospital of Saint-Étienne (CHU Saint-Étienne); SAINBIOSE INSERM U1059, Université Jean Monnet}, \orgaddress{\state{Saint-Étienne}, \country{France}}}

\address[5]{\orgdiv{Département d’Anesthésie-Réanimation}, \orgname{Hôpital Nord, CHU de Saint-Etienne, F-42055}, \orgaddress{\state{Saint-Etienne}, \country{France}}}

\address[6]{\orgdiv{Department of Medicine}, \orgname{Leiden University Medical Center (LUMC)}, \orgaddress{\state{Leiden}, \country{The Netherlands}}}

\address[7]{\orgdiv{Department of Respiratory Medicine and Intensive Care}, \orgname{Université Paris Cité, Hôpital Européen Georges-Pompidou, AP-HP, INSERM UMR-S 970, F-CRIN INNOVTE}, \orgaddress{\state{Paris}, \country{France}}}

\address[8]{\orgdiv{Department of Pneumology}, \orgname{Brest University Hospital (CHU de Brest)}, \orgaddress{\state{Brest}, \country{France}}}

\address[9]{\orgdiv{INSERM U1304 – GETBO (Groupe d’Étude de la Thrombose de Bretagne Occidentale)}, \orgname{Univ Brest, F-CRIN INNOVTE}, \orgaddress{\state{Brest}, \country{France}}}

\corres{Corresponding author Yohan Bard, \email{yohan.bard@inserm.fr}}

\presentaddress{28 avenue Pierre Mendès France,\\ 42 270 Saint-Priest-en-Jarez}


\abstract[Abstract]{Stepped-wedge cluster randomised trials (SW-CRTs) increasingly evaluate complex interventions, yet methodological guidance for analysing composite endpoints using generalized pairwise comparisons (GPC) remains limited. This work investigates the performance of several GPC-based estimators in the presence of clustering, temporal trends, and varying correlation structures typical of SW-CRTs. We conducted an extensive simulation study covering a range of intraclass correlations (ICC), cluster autocorrelation coefficients (CAC), time effects, and treatment effect sizes. Eight analytical approaches were compared, including unadjusted estimators, cluster-stratified win odds, mixed-effects models applied to cluster--period win odds, and probabilistic index models (PIMs). Type~I error control was strongly compromised for methods ignoring time or clustering, whereas only two approaches consistently maintained nominal error rates: a hierarchical mixed-effects model with sequence- and cluster-level random slopes (b4) and a cluster-restricted PIM (c2). These two methods were further evaluated in terms of statistical power, where c2 generally showed higher efficiency, particularly under strong clustering, low CAC, or the presence of temporal trends, while both converged to similar performance for large treatment effects. Overall, our findings identify b4 and c2 as the most reliable GPC-based strategies for SW-CRT analysis and provide practical guidance for their application, including for ongoing trials such as ETHER.
}


\jnlcitation{\cname{%
\author{Taylor M.},
\author{Lauritzen P},
\author{Erath C}, and
\author{Mittal R}}.
\ctitle{On simplifying âincremental remapâ-based transport schemes.} \cjournal{\it J Comput Phys.} \cvol{2021;00(00):1--18}.}

\maketitle

\section{Introduction}\label{sec1}

Generalized Pairwise Comparisons (GPC) is a modern statistical method used to assess the effect of an intervention by considering multiple outcomes simultaneously. This approach addresses the limitations of traditional analyses, such as classical composite endpoints, which assign equal weight to each component event, even when these events differ greatly in clinical importance---for example, fatal versus nonfatal events. As a result, conventional composites may overlook clinically meaningful events that occur less frequently or later than less significant ones \cite{dong2016}.\\

GPC allows investigators to handle composite endpoints where the components are not clinically equivalent by ranking outcomes according to their clinical importance. Intervention effect can then be quantified by clear summary measures, such as the \emph{win ratio}, which is becoming increasingly common, especially in cardiovascular trials \cite{pocock2012}. First introduced by Buyse \cite{buyse2010}, the GPC framework works by comparing all possible patient pairs across treatment groups. For each pair, the outcomes are compared to determine whether one patient had a better (\emph{favorable}), worse (\emph{unfavorable}), or similar (\emph{neutral}) result. The method then aggregates these pairwise results to estimate the treatment effect. More recently, the framework has been expanded to several situations, including dose--response analysis \cite{johns2023}, covariate adjustment using probabilistic index models (PIMs) \cite{pim2012,song2023}, and variable selection strategies \cite{mao2025}, illustrating its flexibility. However, so far in clinical research, GPC has mostly been used in individually randomized trials.\\

In some cases, individual randomization is not feasible, especially when there is a risk that the interventions in different arms might influence each other, a phenomenon known as contamination. In such settings, cluster randomization is often preferred for practical, ethical, or logistical reasons. This approach randomizes groups (or ``clusters'') of participants rather than individuals and can help avoid contamination, simplify trial organization, and improve participant engagement \cite{giraudeau2024}.  

However, applying the win ratio directly in cluster trials presents methodological challenges. Standard formulations of the win ratio assume independent observations, and ignoring the correlation of outcomes within clusters can lead to underestimated variance and overly optimistic confidence intervals, potentially affecting the validity of statistical inference. For instance, Romiti et al.~\cite{romiti2023} applied the generalized pairwise comparison framework in a cluster-randomized setting without explicitly adjusting for intracluster correlation. Their work illustrates how current analytical tools are being used in practice and underscores the need for methods specifically designed for clustered data.\\

When interventions are difficult to implement simultaneously across all clusters, the \emph{stepped-wedge cluster randomized trial} (SW-CRT) offers a practical alternative. In this design, clusters gradually switch from control to intervention over successive periods. This staggered rollout has ethical advantages---all clusters eventually receive the intervention---and facilitates implementation. Variants such as the batched stepped wedge design have also been introduced to accommodate delays in recruitment or intervention roll-out, further increasing the practicality of this design \cite{kasza2022}.\\

However, analyzing data from SW-CRTs poses additional challenges. 
First, individuals within the same cluster tend to have correlated outcomes, which is commonly quantified through the intra-cluster correlation coefficient (ICC) and reflects shared contextual or organizational factors. 
Second, because clusters are observed repeatedly over time, correlations may also persist across periods within the same cluster; this temporal persistence of cluster effects can be characterised by a cluster autocorrelation coefficient (CAC). 
In addition, treatment effects may be confounded by time trends in baseline risk, since clusters initiate the intervention at different times. 
As a result, statistical models must carefully adjust for both clustering and time to avoid biased estimates \cite{billot2024,li2021}. 
Guidelines for the reporting of SW-CRTs have also been published to ensure that the complex features of these designs are clearly described \cite{hemming2018}.\\

Although generalized pairwise comparisons (GPC) are increasingly used in clinical research, their application has so far remained largely confined to individually randomized trials. To date, no methodological work has adapted GPC to explicitly account for clustering or temporal effects in complex randomized designs, such as classical cluster randomized trials or stepped-wedge variants. This paper contributes to addressing this methodological limitation by extending the GPC framework to the stepped-wedge setting, allowing treatment effects to be analysed while properly accounting for both intracluster correlation and secular time trends.

\section{Methods}\label{sec2}

\subsection{Stepped-wedge cluster randomised trial design}

\begin{figure}[h]
	\begin{center}
	\includegraphics[width=0.75\linewidth, scale=0.6]{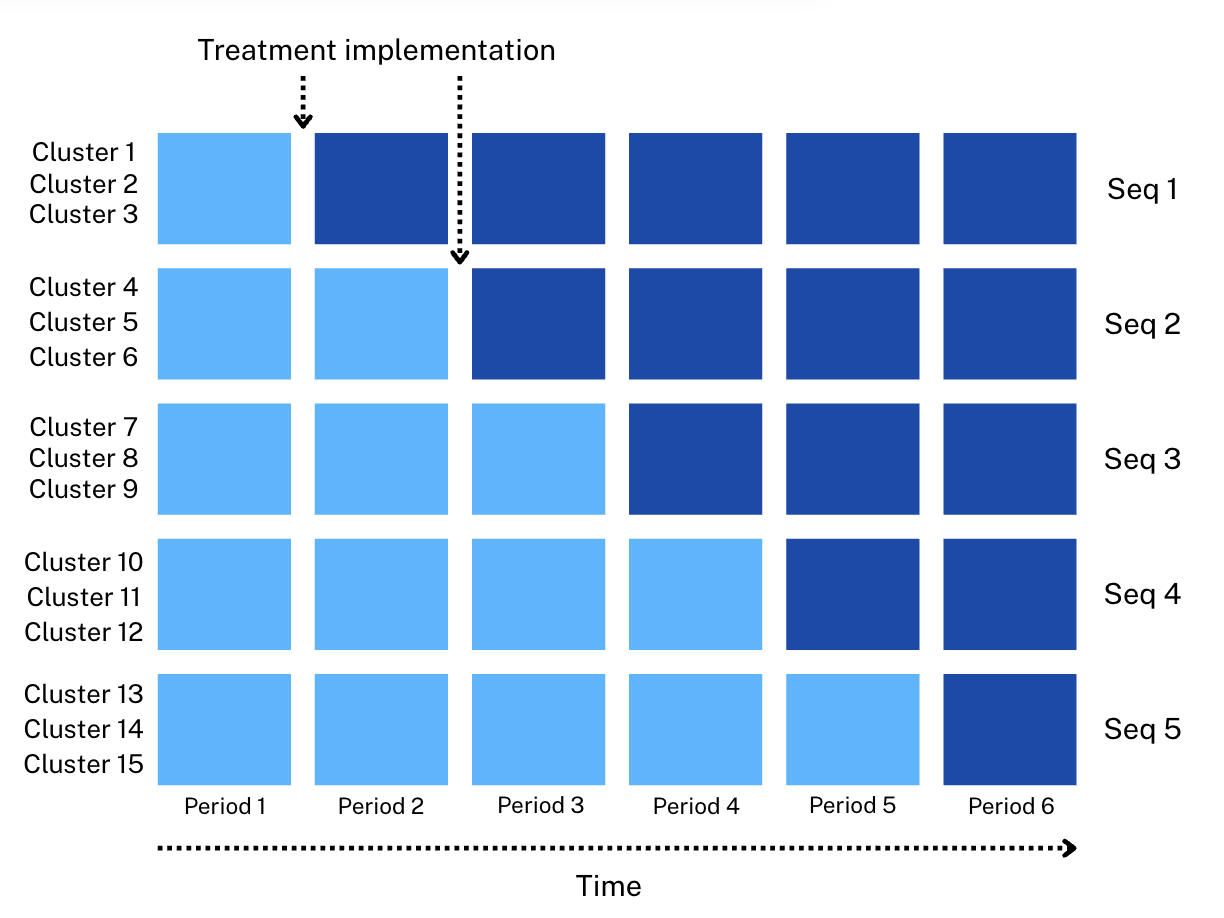}
	\caption{Example of stepped-wedge trial with 5 sequences (6 periods)\label{fig1}}
	\end{center}
\end{figure}

A stepped-wedge cluster randomised trial (SW-CRT) is a design in which clusters cross over from control to intervention at different, pre-specified time periods according to a randomised sequence. The design proceeds in discrete periods, indexed by $j = 1, \dots, J$, during which outcomes are observed. Clusters are indexed by $k = 1, \dots, K$. Within each cluster-period combination $(j, k)$, the number of observed individuals is denoted by $n_{jk}$.The general structure of a stepped-wedge design is illustrated in Figure \ref{fig1}.  \\

Each cluster follows an assigned sequence that determines the period at which it initiates the intervention. By the end of the study, all clusters have received the intervention, ensuring that data are available under both conditions across the trial. The intervention status of a cluster in period $j$ is therefore fully determined by its sequence allocation and the period index.  \\

Although the present work focuses on stepped-wedge designs, the notation and methodological developments introduced throughout this section naturally encompass classical cluster randomised trials as a special case. In particular, a parallel-arm cluster trial corresponds to a stepped-wedge design with a single transition period, so that the proposed framework remains applicable beyond the stepped-wedge setting.\\

Two key sources of dependence arise in SW-CRTs. First, outcomes measured within the same cluster tend to be correlated due to shared clinical practices, patient populations, or organisational factors. This intra-cluster dependence is commonly summarised by the intraclass correlation coefficient (ICC), which quantifies the proportion of outcome variability attributable to between-cluster differences. Second, because clusters are observed repeatedly over time, correlations may persist across periods within the same cluster. In stepped-wedge designs, this temporal persistence is often imperfect and is characterised by a cluster autocorrelation coefficient (CAC), which reflects the degree to which cluster-specific effects remain stable across successive periods.\\

In addition to correlation structures, stepped-wedge trials are intrinsically exposed to secular time trends, whereby the baseline risk of the outcome evolves over calendar time independently of the intervention. Since intervention exposure is systematically confounded with time in SW-CRTs, failure to account for such trends can lead to biased estimation of the treatment effect. These features motivate the need for statistical methods that jointly adjust for clustering, temporal correlation, and time trends, which will be addressed in the following sections.\\

\subsection{Generalized pairwise comparison for composite outcomes}

Generalized pairwise comparisons (GPC) provide a unified framework for evaluating treatment effects based on individual-level outcomes. The approach considers all possible pairs of individuals, one from the treatment group and one from the control group, and determines which individual has the more favorable outcome according to a pre-specified ranking of endpoints.  

Let $Y_i$ denote the outcome for individual $i$. If a single endpoint is considered, $Y_i$ is scalar. In the more general case of multiple outcomes, we define a vector
\[
Y_i = \big(Y_i^{(1)}, \ldots, Y_i^{(M)}\big),
\]
where outcomes are hierarchically ordered according to their clinical relevance: $Y_i^{(1)}$ is the most important endpoint, and $Y_i^{(M)}$ is the least important, often chosen for its discriminative power.  

Treatment allocation is described by the indicator $X_{ijk}$, which equals $1$ if individual $i$ in cluster $k$ at period $j$ is exposed to the intervention, and $0$ otherwise.  

For a comparison between two individuals $i$ and $i'$ assigned to different treatment groups, we define the comparison score $W_{ii'}$. The index of the first outcome that differs is
\[
m^* = \min \Big\{ m \in \{1, \ldots, M\} : Y_i^{(m)} \neq Y_{i'}^{(m)} \Big\}.
\]
If such an $m^*$ exists, then:
\[
W_{ii'} =
\begin{cases}
1, & \text{if } Y_i^{(m^*)} \succ Y_{i'}^{(m^*)}, \text{(favorable pair)}\\[6pt]
-1, & \text{if } Y_i^{(m^*)} \prec Y_{i'}^{(m^*)}, \text{(unfavorable pair)}\\[6pt]
0, & \text{if no such } m^* \text{ exists (neutral pair)}.
\end{cases}
\]

Based on this comparison score, intervention effect can be estimated using different metrics. Let $N_W$ and $N_L$ denote the total numbers of favorable pairs and unfavorable across all pairwise comparisons between treatment and control individuals. The first metric is the \emph{win ratio}, defined as
\[
\text{WR} = \frac{N_W}{N_L}.
\]
Values greater than $1$ indicate that the treatment group achieves more favorable than unfavorable pairs, suggesting a beneficial effect of the intervention.  

Because neutral pairs may occur frequently, a second metric, the \emph{win odds}, modifies the win ratio by allocating half a favorable and half a unfavorable pair to each neutral one. Let $N_T$ denote the number of neutral pairs. The win odds is de fined as
\[
\text{WO} = \frac{N_W + 0.5\,N_T}{N_L + 0.5\,N_T}.
\]

Although the treatment effect in the mixed-effects models is expressed on the $\log(\text{WO})$ scale, this measure admits a simple and intuitive interpretation through its relationship with the net treatment benefit (NTB). 
The NTB is defined as
\[
\text{NTB} = \frac{\text{WO} - 1}{\text{WO} + 1},
\]
and corresponds to the difference in probabilities $(\mathbb{P}_T > \mathbb{P}_C) - (\mathbb{P}_C > \mathbb{P}_T)$, that is, the probability that a randomly selected treated individual has a more favourable outcome than a control individual minus the reverse probability. 
This quantity lies between $-1$ and $1$ and can be interpreted as an absolute treatment effect on the probability scale. 
For small to moderate treatment effects, a first-order approximation yields $\log(\text{WO}) \approx 2 \times \text{NTB}$, which facilitates interpretation of regression coefficients expressed on the log-win-odds scale.\\

While both measures are consistent with the GPC framework, the win odds is generally preferred in settings with non-negligible numbers of neutral pairs. By incorporating neutral pairs into the calculation, it provides a smoother and more robust estimate of treatment effect, while retaining the same interpretability: values greater than $1$ suggest superiority of the treatment group. Moreover, the win odds is linked to more classical intervention effect estimates. For example, in the special case where there is only one outcome, the win odds could be interpreted as an inverse of a hazard ratio\cite{Oakes2016_WinRatio}. For these reasons, the win odds will be the main effect measure considered in this paper.  \\

\subsection{Covariates adjustment within GPC}

In stepped-wedge cluster randomized trials, potential confounding may arise from several sources: clustering of outcomes within groups, secular time trends, and cluster- or period-level characteristics. A major motivation for this work is that generalized pairwise comparisons may produce biased estimates of the treatment effect when applied without adjustment on these potential confounding sources. To address these issues, we denote by $X$ the set of covariates to be considered, which may include both individual-level and cluster-level variables. In the GPC framework, covariates may be taken into account by different approaches.  

\subsubsection{Stratification}  
The simplest adjustment strategy consists in stratifying pairwise comparisons according to a categorical variable $S = S(X)$. For example, $S$ may represent the cluster or the trial period. Within each stratum $s$, only comparisons between individuals belonging to that stratum are considered, and a stratum-specific win odds $\text{WO}_s$ is estimated.  

These stratum-specific estimates are then combined into a global measure by weighting their logarithms\cite{Dong2023}\cite{Buyse2025}:  
\[
\log(\text{WO}) = \sum_{s} w_s \, \log(\text{WO}_s),
\]
with weights $w_s$ satisfying $\sum_s w_s = 1$. Different weighting schemes may be used, such as inverse-variance weights or weights proportional to the size of each stratum. This stratified approach parallels classical stratified tests, while preserving the win-based interpretation of the effect measure.  

\subsubsection{Regression framework with probabilistic index models}  
To adjust GPC analysis on a potential continuous or categorical covariates $X$, probabilistic index models (PIMs) has been proposed,  embedding GPC into a regression framework. The key idea is that the win odds can be expressed as a transformation of the probabilistic index, defined as
\[
\mathbb{P}\big( Y \succ Y_{\star} \mid X , X_{\star}  \big) = \mathbb{P}\big( Y > Y_{\star} \mid X , X_{\star}  \big) + 0.5 \, \mathbb{P}\big( Y = Y_{\star} \mid X , X_{\star}  \big),
\]
that is, the probability that a randomly selected individual has a more favorable outcome than another randomly selected individual depending on their respective covariates $X$ and $X_{\star}$.

In PIMs, this probability is linked to covariates through a regression model:
\[
g \left( \mathbb{P}\big( Y \succ Y_{\star} \mid X , X_{\star}  \big) \right) =  Z^{T}\beta,
\]
where $g(\cdot)$ is a link function (logit, probit, identity). Vector $Z$ denotes a vector with elements depending on $X$ and $X_{\star}$ like $Z = X - X_{\star}$ for example. The parameter vector $\beta$ contains the regression coefficients describing how each covariate affects the probability of a favorable comparison; in particular, the treatment coefficient corresponds directly to the adjusted log win odds.\\
 For the analysis of SW-CRT, this pairwise regression formulation allows direct adjustment for both individual and cluster-level covariates, as well as time effects, while retaining the pairwise comparison principle. \\ 

Estimation can be performed through standard estimating equations \cite{Thas2012}. The fitted model provides an adjusted estimate of the treatment effect in terms of win odds, together with confidence intervals and $p$-values. Compared to stratification, this regression approach allow to handle directly continuous and/or multiple covariates.

\subsection{Evaluated methods for the analysis of SW-CRT based on GPC}
\label{sec:methods}

In this section, we describe the candidate approaches for analyzing data from stepped-wedge cluster randomized trials (SW-CRTs) using the generalized pairwise comparison (GPC) framework. Each approach provides an estimate of the win odds (WO) while addressing, to varying degrees, the confounding introduced by time dependency and data clustering.

\subsubsection{Methods without adjustment for time trends}

The first class of approaches evaluated deliberately ignores the temporal structure of the stepped-wedge design. The comparison is treated as if the trial were a parallel-group design, with all observations pooled according to treatment status.  

The first approach, referred to as \textbf{Method a1: Crude win odds}, compares all individuals who received the intervention, regardless of period or cluster, against all individuals who remained under control. All possible treatment--control pairs are generated, and the overall win odds is estimated as
\[
\text{WO} = \frac{N_W + 0.5N_T}{N_L + 0.5N_T},
\]
where $N_W$, $N_L$, and $N_T$ denote the total numbers of favorable, unfavorable, and neutral pairs, respectively. This approach is simple but ignores both intra-cluster correlation and time dependency effects.  

A refinement, denoted \textbf{Method a2: Cluster-stratified win odds}, accounts for clustering by computing the win odds within each cluster separately. In each cluster $k$, individuals under control are compared to those under intervention, producing a cluster-specific win odds $\text{WO}_k$. These estimates are then aggregated across clusters to obtain an overall measure. Aggregation is performed on the log scale to stabilise the variance, with the global estimate given by
\[
\log(\text{WO}) = \sum_{k=1}^K w_k \, \log(\text{WO}_k),
\]
where $w_k$ are non-negative weights satisfying $\sum_{k=1}^K w_k = 1$. \\

Let $m_k$ and $n_k$ denote the sample sizes in the control and intervention arm, respectively, for cluster $k$. 
The total number of treatment--control pairs that can be formed within this cluster is
\[
M_k = m_k \times n_k.
\]

A natural choice is to let $w_k$ be proportional to $M_k$, i.e.
\[
w_k = \frac{M_k}{\sum_{\ell=1}^K M_\ell}.
\]

Here, $M_k$ represents the total number of available treatment--control comparisons in cluster $k$, independently of whether these pairs later yield favorable, unfavorable, or tied outcomes.

This weighting scheme \cite{buyse2010}\cite{Buyse2025} ensures that clusters contributing more information have greater influence on the overall estimate, while avoiding instability that may arise if inverse-variance weights are used (since variances can sometimes be zero when all comparisons within a cluster yield the same outcome). 

Methods (a1) and (a2) are easy to implement and computationally efficient. However, because they do not adjust for time trends, they may be biased in the context of SW-CRTs, where intervention exposure is confounded with calendar time.

\subsubsection{Between-period cluster-level analysis}

To jointly account for clustering and temporal dynamics, a second approach constructs win odds at the level of cluster--period pairs. Specifically, for each cluster $k \in \{1, \dots, K\}$ and for every pair of periods $(j_1, j_2)$ with $j_1 < j_2$, all possible pairwise comparisons are formed between individuals observed in cluster $k$ during period $j_1$ and those observed in the same cluster during period $j_2$. Within this set of comparisons, favorable, unfavorable, and neutral outcomes are classified according to the generalized pairwise comparisons (GPC) framework introduced earlier. Let $N_{W_{k,(j_1,j_2)}}$, $N_{L_{k,(j_1,j_2)}}$, and $N_{T_{k,(j_1,j_2)}}$ denote, respectively, the numbers of favorable, unfavorable, and neutral pairs obtained from these within-cluster comparisons.

The win odds for cluster $k$ between periods $(j_1, j_2)$ are then defined as
\[
\text{WO}_{k,(j_1,j_2)} = \frac{N_{W_{k,(j_1,j_2)}} + 0.5\, N_{T_{k,(j_1,j_2)}}}{N_{L_{k,(j_1,j_2)}} + 0.5\, N_{T_{k,(j_1,j_2)}}}.
\]

Because each cluster provides several estimates of $\text{WO}_{k,(j_1,j_2)}$ across different period pairs, these estimates must subsequently be pooled to obtain an overall intervention effect while accounting for time trends and between-cluster variability. To this end, the logarithm of the cluster--period win odds, $\log(\text{WO}_{k,(j_1,j_2)})$, is modeled using a mixed-effects model that includes fixed effects for time and intervention, and random effects to capture residual heterogeneity across clusters and periods.\\

We first considered a baseline specification, referred to as \textbf{Model b1}, which includes random intercepts for both clusters and pairs of periods. This model accounts for unobserved heterogeneity across clusters as well as potential correlation between comparisons involving the same pair of periods:
\[
\log \big( \text{WO}_{k,(j_1,j_2)} \big) = \alpha + u_k + v_{(j_1,j_2)} + \delta \times (X_{k,j_2}-X_{k,j_1}) + \gamma \times (j_2 - j_1) + \varepsilon_{k,(j_1,j_2)},
\]
where $u_k \sim \mathcal{N}(0,\sigma^2_C)$ is a random intercept for cluster $k$, and $v_{(j_1,j_2)} \sim \mathcal{N}(0,\sigma^2_P)$ is an additional random intercept capturing variability specific to the period pair $(j_1,j_2)$.  \\

While this structure captures baseline heterogeneity across clusters, it assumes that the intervention effect is constant across all clusters. To relax this assumption, \textbf{Model b2} introduces a random slope for the treatment difference at the cluster level. This allows each cluster to have its own deviation from the overall treatment effect, acknowledging that contextual or implementation factors may cause the effect of the intervention to vary across clusters:
\[
\log \big( \text{WO}_{k,(j_1,j_2)} \big) = \alpha + u_k + v_{(j_1,j_2)} + (\delta + z_k) \times (X_{k,j_2}-X_{k,j_1}) + \gamma \times (j_2 - j_1) + \varepsilon_{k,(j_1,j_2)},
\]
with $z_k \sim \mathcal{N}(0,\sigma^2_{Int})$ representing the random deviation of the treatment effect for cluster $k$.  \\

Although Model b2 allows for cluster-specific treatment effects, it does not explicitly account for the stepped-wedge structure, where clusters are organized into sequences of crossover transitions. To reflect potential differences in treatment effects between these sequences, \textbf{Model b3} introduces a random slope at the sequence level. This specification recognizes that clusters belonging to the same sequence share similar exposure patterns and may thus exhibit correlated responses to the intervention:
\[
\log \big( \text{WO}_{k,(j_1,j_2)} \big) = \alpha + u_k + v_{(j_1,j_2)} + (\delta + w_{s(k)}) \times (X_{k,j_2}-X_{k,j_1}) + \gamma \times (j_2 - j_1) + \varepsilon_{k,(j_1,j_2)},
\]
where $w_{s(k)} \sim \mathcal{N}(0,\sigma^2_S)$ captures the random slope for the treatment difference associated with the sequence $s(k)$ to which cluster $k$ belongs.  \\

Building upon this idea, \textbf{Model b4} proposes a more flexible hierarchical specification that nests cluster-level random slopes within sequences. This model captures both the shared variability of treatment effects across sequences and the residual heterogeneity between clusters within the same sequence:
\[
\log \big( \text{WO}_{k,(j_1,j_2)} \big)
= \alpha + u_k + v_{(j_1,j_2)}
+ (\delta + w_{s(k)} + z_k) \times (X_{k,j_2} - X_{k,j_1})
+ \gamma \times (j_2 - j_1)
+ \varepsilon_{k,(j_1,j_2)},
\]
where $w_{s(k)} \sim \mathcal{N}(0, \sigma^2_S)$ a sequence-level random slope, and $z_k \sim \mathcal{N}(0, \sigma^2_{C|S})$ a cluster-level deviation nested within the same sequence.This structure allows the model to capture both large-scale differences between sequences and fine-scale heterogeneity among clusters.\\ 

Because the number of clusters is often limited in SW-CRTs, standard variance estimators and asymptotic approximations may yield biased inference. The Kenward--Roger correction \cite{KenwardRoger1997} was applied systematically  to correct small-sample bias and Type I error rates.

\subsubsection{Individual-level analysis}

A third strategy is to analyse the data directly at the level of individual pairwise comparisons, using the probabilistic index model (PIM) framework. In this setting, each observation corresponds to an individual $i$ from cluster $k$ observed at period $j$, with outcome $Y_{ijk}$, treatment indicator $X_{ijk}$, and observation time $t_{ijk}$. The PIM approach allows modelling the probability that one individual has a more favorable outcome than another, given their respective covariates, thereby embedding the concept of win odds into a regression framework \cite{pim2012}.  \\

In the simplest specification, referred to as \textbf{Model c1 (Standard PIM)}, let $(i,j,k)$ and $(i',j',k')$ denote two individuals drawn from the trial. The model is defined as
\[
g\left( \mathbb{P}\big( Y_{ijk} \succ Y_{i'j'k'} \,\big|\, X_{ijk},X_{i'j'k'},t_{ijk},t_{i'j'k'} \big) \right) 
= \alpha + \delta \times (X_{ijk} - X_{i'j'k'}) + \gamma \times (t_{ijk} - t_{i'j'k'}),
\]
where $g(\cdot)$ is a link function, typically the logit link.  \\

In this formulation, $\delta$ quantifies the adjusted treatment effect in terms of win odds, while $\gamma$ captures a linear time trend at the individual level. All possible pairs of individuals are considered, including treatment--treatment, control--control, and treatment--control pairs. Comparisons within the same treatment arm contribute to the estimation of the temporal effect, whereas cross-arm comparisons inform both treatment and time effects. This formulation is flexible and efficient because it uses all available information in the dataset.  \\

However, standard PIM estimation assumes independence between observations or, at least, robustness to mild dependence. In stepped-wedge cluster randomised trials, the hierarchical structure of the data induces strong intra-cluster correlation, as individuals within the same cluster are exposed to shared contextual and temporal factors. Although the PIM framework is generally robust to moderate correlation structures, ignoring this clustering can lead to underestimated variance and inflated Type I error.  \\

To address this issue, an alternative specification called \textbf{Model c2 (Cluster-restricted PIM)} limits the comparisons to individuals belonging to the same cluster, i.e. $k = k'$. The model then becomes
\[
g\left( \mathbb{P}\big( Y_{ijk} \succ Y_{i'j'k} \,\big|\, X_{ijk},X_{i'j'k},t_{ijk},t_{i'j'k} \big) \right) 
= \alpha + \delta \times (X_{ijk} - X_{i'j'k}) + \gamma \times (t_{ijk} - t_{i'j'k}).
\]

By restricting comparisons within clusters, this approach eliminates between-cluster heterogeneity and ensures that the estimated treatment effect is based solely on within-cluster contrasts. This structure directly aligns with the randomisation mechanism of SW-CRTs, where treatment allocation occurs at the cluster level, and helps prevent bias that could arise from cross-cluster comparisons.  \\

Overall, both PIM specifications provide an adjusted estimate of the treatment effect expressed on the win-odds scale. The standard PIM (c1) leverages all available treatment--control pairs, which can increase efficiency but may be more sensitive to cluster-level correlation. In contrast, the cluster-restricted PIM (c2) offers a more conservative yet robust approach, better respecting the stepped-wedge design hierarchical structure and mitigating the influence of intra-cluster dependencies.

\subsection{Simulation study}
\label{sec:simulation-study}

To evaluate the statistical performance of the proposed methods, we conducted a simulation study based on a stepped-wedge cluster randomised design. The objective was to assess how the different analytical approaches behave under varying degrees of clustering, temporal correlation, and treatment effect, while maintaining consistency with the structure of typical SW-CRTs.\\

We considered a design with $K=45$ clusters, distributed across 5 sequences, with clusters sequentially transitioning from control to intervention over $J=6$ periods. In each cluster and period, $n=10$ individuals were observed. These values were choosen has it corresponds to the characteristics of the ETHER trial discussed below. A continuous observation time $t_{ijk}$ was assigned to each individual $i$ in cluster $k$ at period $j$, sampled uniformly within the corresponding period time interval.\\

The individual binary outcome $Y_{ijk}$ was generated from a Bernoulli distribution with success probability $\pi_{ijk}$:
\[
Y_{ijk} \sim \text{Bernoulli}(\pi_{ijk}),
\]
and the success probability followed a logistic model of the form
\[
\text{logit}(\pi_{ijk}) = \log\!\left(\frac{p_0}{1-p_0}\right) + \delta \cdot X_{ijk} + \eta_k + \zeta_{jk} + \beta_t \, t_{ijk}.
\]
Here, $p_0$ denotes the baseline probability under control, $X_{ijk}$ is the treatment indicator ($X_{ijk}=1$ if the individual is exposed to the intervention, $0$ otherwise), and $\beta_t$ represents a continuous time effect modelling secular trends.  \\

The random effects $\eta_k$ and $\zeta_{jk}$ were introduced to capture within-cluster dependencies. The first term, $\eta_k \sim \mathcal{N}(0, \sigma^2_{\text{cluster}})$, represents a persistent cluster-level deviation that induces correlation between all individuals from the same cluster. The second term, $\zeta_{jk} \sim \mathcal{N}(0, \sigma^2_{\text{period}})$, accounts for between period variation specific to a given cluster, thereby introducing additional dependence between individuals observed within the same period. Both random effects were assumed mutually independent.\\

Under this formulation, the total correlation between two individuals from the same cluster can be expressed through the intraclass correlation coefficient (ICC):
\[
\text{ICC} = \frac{\sigma^2_{\text{cluster}} + \sigma^2_{\text{period}}}{\sigma^2_{\text{cluster}} + \sigma^2_{\text{period}} + \pi^2/3},
\]
where $\pi^2/3$ is the variance of the logistic distribution\cite{Morgan2017}. The ICC thus quantifies the overall proportion of variance attributable to the shared cluster-level effects.  \\

In stepped-wedge trials, the same clusters are repeatedly observed across periods, the correlation structure also vary over time. This temporal dependence can be quantified by the cluster autocorrelation coefficient (CAC), defined as the proportion of the ICC that remains constant across periods within a given cluster. The CAC thus controls the persistence of the cluster effect over time :  \\
\[
\text{CAC} = \frac{ \sigma^2_{\text{cluster}}  }{\sigma^2_{\text{cluster}} + \sigma^2_{\text{period}}},
\]
Within the simulation study, the variances of each random effect was calculated from predefined ICC and CAC values\cite{Hooper2016} in the following way :
\[
\sigma^2_{\text{cluster}} = \text{CAC} \times \text{ICC} \times \frac{\pi^2/3}{1 - \text{ICC}}, \qquad
\sigma^2_{\text{period}} = (1 - \text{CAC}) \times \text{ICC} \times \frac{\pi^2/3}{1 - \text{ICC}}.
\]
This parameterization ensures that the correlation between two individuals from the same cluster and period equals $\text{ICC}$, while the correlation between two individuals from the same cluster but different periods equals $\text{CAC} \times \text{ICC}$. When $\text{CAC}=1$, the cluster effect is perfectly stable across time, corresponding to the standard cross-sectional ICC. As $\text{CAC}$ decreases, the cluster effect becomes more variable across periods, reflecting a gradual loss of temporal correlation within clusters.\\

Each simulation scenario was defined by a unique combination of parameters controlling baseline risk, treatment effect, correlation structure, and time trend. Specifically, we varied the baseline event probability $p_0 \in \{0.01, 0.1, 0.2, 0.5\}$, the intraclass correlation $\text{ICC} \in \{0, 0.05, 0.1, 0.3\}$, the cluster autocorrelation $\text{CAC} \in \{0.5, 0.75, 0.9, 1\}$, and the time effect $\beta_t \in \{-0.05, -0.025, 0, 0.025, 0.05\}$. The treatment effect was specified through the log-odds ratio parameter $\delta \in \{0, 0.25, 0.5, 1, 1.5\}$. For each configuration, 500 independent datasets were simulated to ensure reliable estimation of empirical performance.\\

Each simulated dataset was then analysed using all candidate methods described previously: (i) win odds without time adjustment (crude and cluster-stratified), (ii) cluster-level analyses using linear mixed-effects models, and (iii) individual-level pairwise comparison approaches based on probabilistic index models (PIMs).  \\

The win odds were computed using the \texttt{BuyseTest} package in \texttt{R}, probabilistic index models were fitted with the \texttt{pim} package, and mixed-effects models were estimated using \texttt{lme4} and \texttt{lmerTest}. For each method, we extracted the estimated win odds, its standard error, and the corresponding $p$-value. Scenarios with $\delta=0$ were used to evaluate type I error, while non-null configurations allowed assessment of statistical power.\\

\subsection{Power analysis of the ETHER trial}

The ETHER trial (\textit{Empowering paTiEnts for sHared dEcision in anticoagulant theRapy}, NCT06731244) is part of the European research program MORPHEUS (Grant HORIZON-HLTH-2022-11-01). Its objective is to improve the management of patients with unprovoked venous thromboembolism (VTE) through aa shared decision making strategy integrating multicomponent risk prediction scores and socio-anthropological scales (TDMI). The TDMI is designed to be used during clinical consultations and provides patients and clinicians with individualized, evidence-based information on the benefits and risks of long-term anticoagulation, taking into account patient characteristics, clinical history, and preferences. The trial evaluates whether the TDMI improves decision-sharing, treatment safety, efficiency and patient engagement during long-term anticoagulation.\\

ETHER was designed as a stepped-wedge cluster randomised trial including 45 clinical centers distributed over 5 sequences and 6 time periods. The first period lasts 12 months, followed by five successive 3-month periods during which clusters sequentially transition from control to intervention. \\

The primary endpoint of ETHER is a composite endpoint composed of the following ordered criteria:
\begin{enumerate}
  \item all-cause mortality,
  \item symptomatic venous thromboembolism (VTE) recurrence,
  \item major or clinically relevant non-major bleeding,
  \item patient activation measured using the PAM-13 score \cite{hibbard2004pam13}.
\end{enumerate}

The original sample size was determined according to the protocol assumptions by first calibrating the detectable treatment effects for a classical parallel-group randomised trial, and then applying an appropriate design effect (scaling factor) to account for the stepped-wedge structure. 
Because this procedure was not specifically tailored to GPC-based estimators, the present simulation assesses the operating characteristics (power and Type-I error) of the GPC methods under the fixed sample size defined in the protocol.\\


The simulation framework reproduced the stepped-wedge structure of the ETHER trial, with $K = 45$ clusters observed over six time periods. For each cluster-period, $n = 10$ patients were simulated, each with a continuous observation time $t_{ijk}$ uniformly distributed within the corresponding period.\\

For the three binary endpoints (mortality, VTE recurrence, and bleeding), outcomes were generated according to the logistic model described in Section~\ref{sec:simulation-study}, using the same simulation procedure. The baseline event probabilities ($p_0$) and treatment effects ($\delta$) for each criterion were set according to protocol assumptions and are summarised in Table~\ref{tab:params}.
\\

\begin{table}
\centering
\caption{Baseline probabilities ($p_0$) and treatment effects ($\delta$) used for the binary outcomes in the ETHER simulation.}
\label{tab:params}
\begin{tabular}{lcc}
\toprule
\textbf{Criterion} & \textbf{Baseline risk $p_0$} & \textbf{Treatment effect $\delta$ (Odds Ratio)} \\
\midrule
All-cause mortality & 0.10 & $-0.117$ (0.89) \\
VTE recurrence & 0.066 & $-0.674$ (0.51) \\
Major or CRNM bleeding & 0.070 & $-0.816$ (0.44)\\
\bottomrule
\end{tabular}
\end{table}

For the continuous endpoint, the PAM-13 score was simulated as:
\begin{equation*}
\text{PAM}_{ijk} \sim \mathcal{N}(\mu_{ijk}, \sigma^2_{\text{PAM}}),
\qquad
\mu_{ijk} = 62.6 + \delta_{\text{PAM}} X_{ijk} + u_k + v_{jk},
\label{eq:pam}
\end{equation*}
with $\delta_{\text{PAM}} = 5$ (expected mean increase) and $\sigma_{\text{PAM}} = 13.6$ \cite{wilkinson2025}. Random effects $u_k$ and $v_{jk}$ representing between cluster and between period variability. Their respective variance was calculated based on predefined   ICC and CAC values as in the binary models by replacing  $\pi^2/3$ by $\sigma^2_{\text{PAM}}$. Simulated scores were truncated to the admissible range $[0, 100]$. In hierarchical pairwise comparisons, absolute PAM differences smaller than 5.4 points were considered ties, reflecting clinically negligible variation \cite{wilkinson2025}.
\\[0.5em]


For each simulated dataset, hierarchical pairwise comparisons were constructed: two individuals were compared sequentially according to the ordered list of outcomes above. The first non-tied criterion determined a favorable, unfavorable, or neutral, following the GPC framework. This procedure yielded, for each cluster and period pair, counts of favourable, unfavourable, and tied comparisons.\\

Cluster--period win odds were then computed and modelled using the hierarchical mixed-effects model described in Section~\ref{sec:methods} (Model b4). In parallel, individual-level comparisons were analysed using the within-cluster probabilistic index model (Model c2). These two models were selected because the simulation study identified them as the only approaches providing adequate control of type~I error while maintaining competitive statistical power. Both models were applied sequentially as additional hierarchical outcomes were incorporated.\\

Simulation scenarios explored combinations of $\text{ICC}\in\{0.01,0.03,0.05,0.10\}$, $\text{CAC}\in\{0.5,0.75,0.9,1.0\}$, and $\beta_t\in\{-0.05,-0.025,0,0.025,0.05\}$. Each configuration was replicated 500 times to evaluate Type-I error rate, and statistical power under the fixed sample size defined by the ETHER protocol.

\section{Results}

\subsection{Simulation study}

\subsubsection{Type I error}

\begin{figure}[ht]
    \centering
    \includegraphics[width=\textwidth]{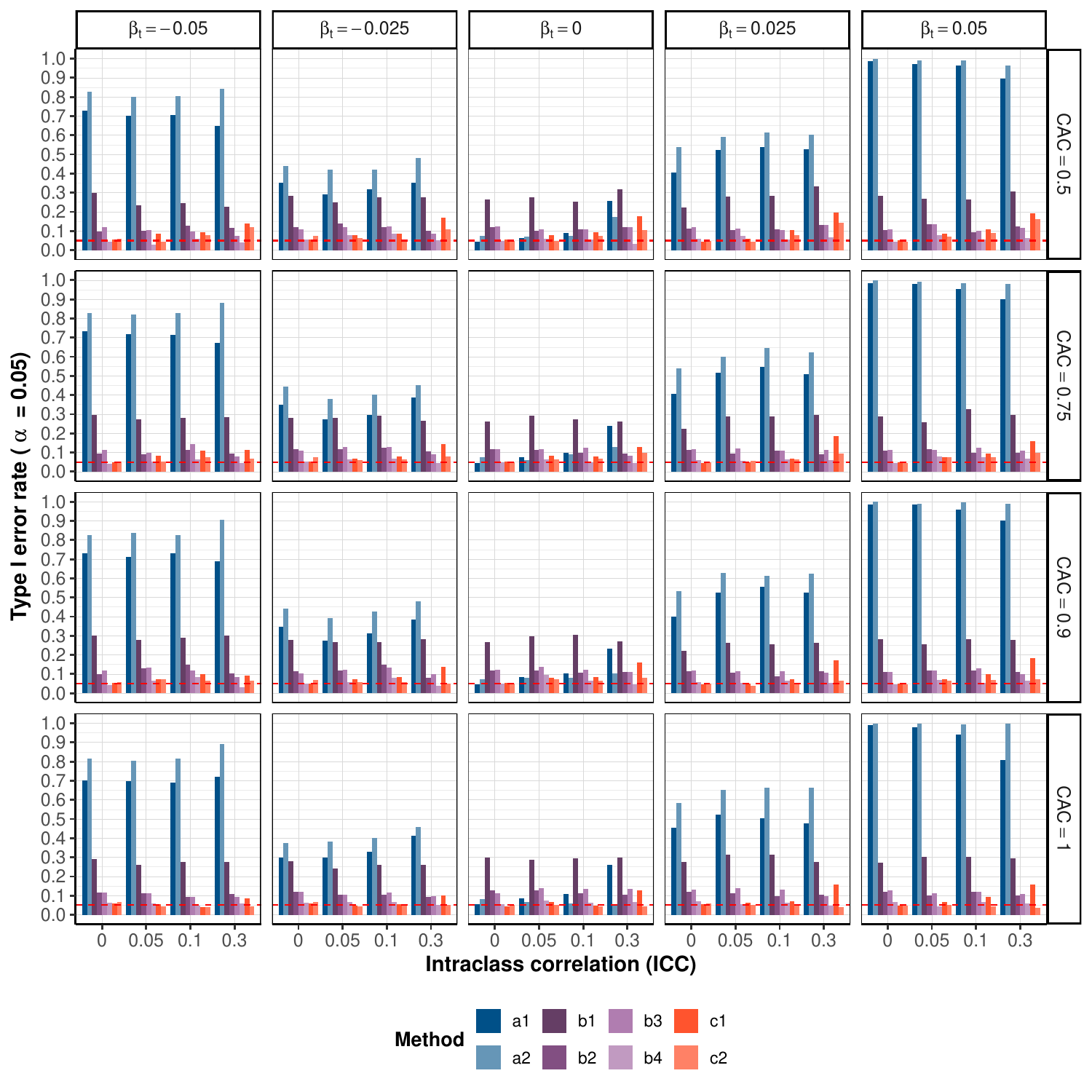}
    \caption{
    Empirical type I error rates ($\alpha = 0.05$) for all evaluated methods under the null hypothesis of no treatment effect ($\delta=0$), for baseline risk $p_0 = 0.1$. 
    The results are shown as a function of the intraclass correlation (ICC), and stratified by the cluster autocorrelation coefficient (CAC) and the secular time trend ($\beta_t$). 
    Methods a1--a2 (unadjusted), b1--b4 (cluster-level mixed models), and c1--c2 (probabilistic index models) are compared.
    The dashed red line indicates the nominal 5\% level.
    }
    \label{fig:FP_all_methods}
\end{figure}

We first evaluated the empirical type I error rate under the null hypothesis of no treatment effect ($\delta = 0$). 
Figure~\ref{fig:FP_all_methods} summarizes the results for all candidate methods across combinations of ICC, CAC, and secular time trend $\beta_t$ (with $p_0 = 0.2$).\\

When there was no time trend ($\beta_t = 0$) and no correlation between cluster (ICC = 0), both unadjusted methods (a1--a2) maintained the nominal 5\% error rate. 
However, as soon as between-cluster variability increased, the crude estimator (a1) exhibited progressively inflated type I error, with rejection rates rising proportionally to the ICC. 
The cluster-stratified estimator (a2) maintained the nominal 5\% error rate in case of no or small between period variability (CAC=1 or 0.9) but fails with CAC values of 0.75 or 0.5.
In the presence of any secular trend ($\beta_t \neq 0$), neither a1 nor a2 provided valid inference, confirming that ignoring time leads to structural confounding in stepped-wedge designs.\\

Among cluster-level mixed-effects models, the simplest formulations (b1--b2) showed substantial inflation across nearly all scenarios, often exceeding 20--30\% even under mild clustering or weak time trends. 
Including a sequence-level random slope (b3) mitigated but did not eliminate the inflation. \\
Only the fully hierarchical specification incorporating both sequence and cluster--level random slopes (b4) controlled the type I error consistently, remaining close to the nominal level across all ICC, CAC and $\beta_t$ combinations.\\

For individual-level approaches, the standard PIM (c1) behaved correctly for ICC values below  0.1; its type I error increased markedly  for greater ICC values. 
The cluster-restricted PIM (c2) generally maintained correct type I error across the range of scenarios, in case with no or small between period variability (CAC=1 or 0.9). For case with larger between period variability (CAC values of 0.75 or 0.5), c2 tended to show a moderate inflation but with rates under 10\%. 
This inflation became more pronounced in case with large between cluster and period variabilities (ICC=0.3 and CAC = 0.5) with type I error rates of 15\%.\\

Overall, Figure~\ref{fig:FP_all_methods} clearly identifies b4 and c2 as the only methods offering reliable control of the type I error in stepped-wedge GPC analyses.\\

\subsubsection{Statistical power}

We next evaluated statistical power under non-null treatment effects ($\delta > 0$), focusing exclusively on the two methods that provided acceptable control of the type I error:  
the hierarchical mixed-effects model with sequence- and cluster-level random slopes (b4),  
and the cluster-restricted probabilistic index model (c2).  

\begin{figure}[h]
\centering
\includegraphics[width=\textwidth]{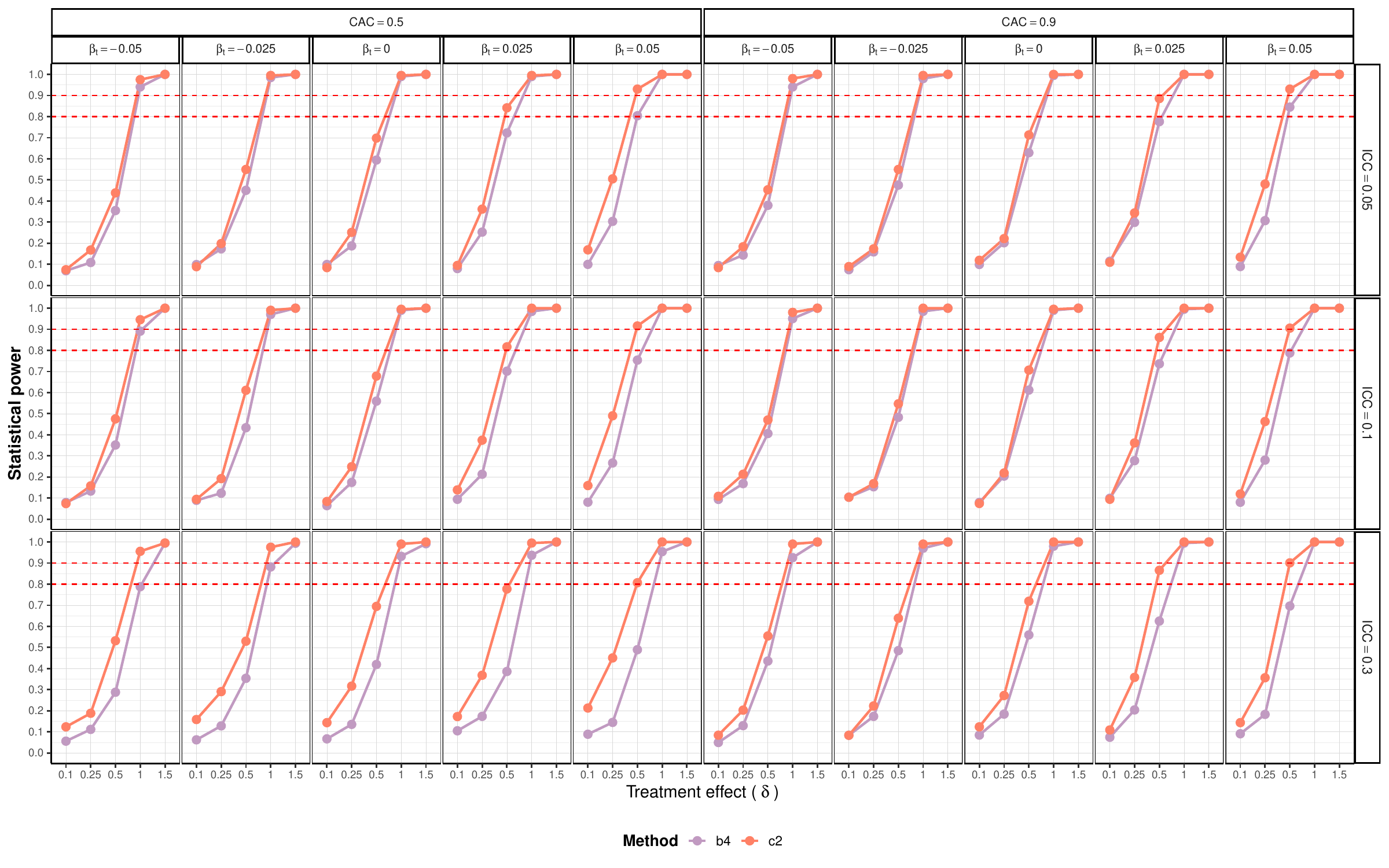}
\caption{
Statistical power of the two best-performing methods, the hierarchical mixed-effects model with sequence- and cluster-level random slopes (Model b4) and the cluster-restricted probabilistic index model (Model c2), across all non-null treatment effect scenarios ($\delta > 0$) for baseline risk $p_0=0.1$. 
For each value of the treatment effect ($\delta$), two bars are shown side-by-side corresponding to the two methods. 
Panels are stratified by intraclass correlation (ICC), cluster autocorrelation coefficient (CAC), and the secular time trend $\beta_t$. 
The horizontal dashed lines at 80\% and 90\% indicate a commonly used power benchmark.}
\label{fig:power_ratioP_methods}
\end{figure}

Power was assessed across increasing values of the treatment effect parameter ($\delta$), and stratified by ICC, CAC, and the secular time trend $\beta_t$.  
Results for all combinations of these parameters are summarized in Figure~\ref{fig:power_ratioP_methods}.  
For readability, the figure displays a representative subset of ICC--CAC combinations, while the full set of power plots covering all correlation structures and time-trend scenarios is provided in the Supplementary Material.

Across all scenarios without a secular trend ($\beta_t = 0$), c2 consistently achieved higher power than b4 for small to intermediate treatment effects.  
This advantage became more pronounced as the intraclass correlation (ICC) increased and as the cluster autocorrelation (CAC) decreased, showing that c2 is more robust when within-cluster dependence is strong while period-to-period stability is weak.  
For large treatment effects ($\delta > 1$), both methods converged to similar power, indicating that differences between b4 and c2 are mainly relevant for intermediate effect sizes where separating treatment effects from cluster-level heterogeneity is most challenging.\\

When a positive time trend was introduced ($\beta_t > 0$), the performance gap between the two methods widened further.  
Temporal trends amplified the relative advantage of c2, which achieved noticeably higher power than b4 across nearly all combinations of ICC and CAC.  
The same qualitative patterns persisted:  
the advantage of c2 increased with larger ICC and lower CAC, while both methods again displayed similar performance for large treatment effects.  
These results highlight that temporal structure tends to reinforce the efficiency gains of c2, especially in correlation settings that are difficult for cluster-level models.\\

Overall, both b4 and c2 demonstrated satisfactory statistical performance, with b4 performing competitively in scenarios with weak clustering and strong period-to-period autocorrelation, and c2 providing consistently higher efficiency in all other situations, particularly when temporal trends and complex correlation structures are present.\\

\subsection{Statistical power of the Ether trial with the candidate methods}

\begin{figure}[h]
	\begin{center}
	\includegraphics[width=1\linewidth, scale=1]{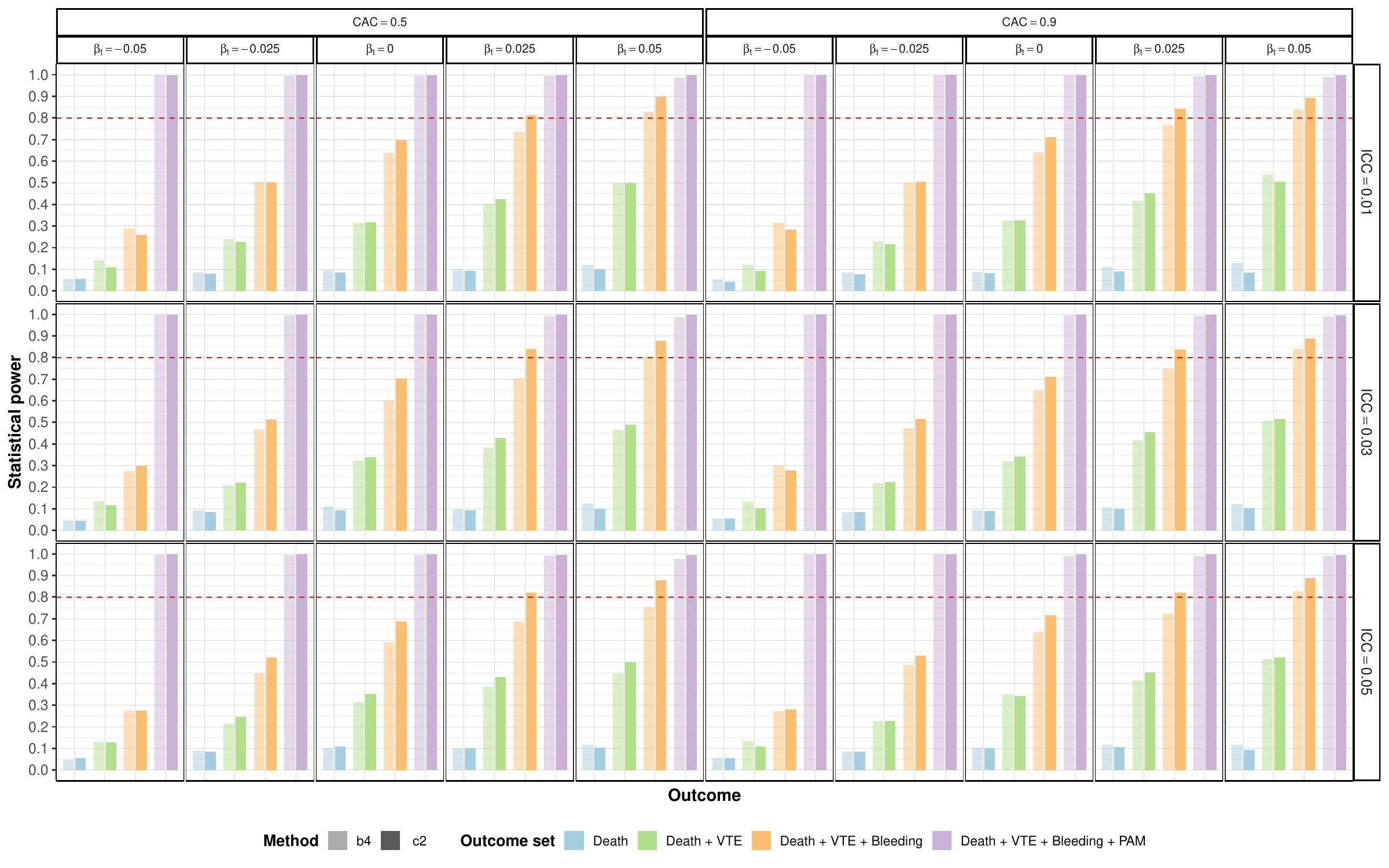}
	\caption{Statistical power of the different methods used to estimate the win odds across the simulation scenarios. Bars represent the proportion of replications in which the null hypothesis is rejected (p < 0.05). Results are stratified by ICC (rows) and by CAC crossed with the temporal effect $\beta_t$ (columns). Pastel colors distinguish the successive outcome sets, while transparency encodes the statistical method: \textbf{Model c2} and \textbf{Model b4}. The horizontal red dashed line indicates the 80\% power threshold.)\label{fig4}}
	\end{center}
\end{figure}

Across all simulation scenarios, statistical power varied systematically according to the number of criteria included in the composite endpoint, as well as the intra-cluster correlation (ICC), the cluster autocorrelation (CAC), and the temporal trend $\beta_t$. Figure~\ref{fig4} summarizes the rejection rates of the null hypothesis (p < 0.05) for each parameter combination.\\

A clear monotonic relationship emerged between the richness of the composite endpoint and the resulting power. When only death was used as the endpoint, power remained very low (approximately 5\%--15\%). Adding VTE increased power substantially (10\%--50\%). Including bleeding as a third component further strengthened performance, with values ranging from about 25\% to 90\%. Finally, when all four outcomes were combined (death + VTE + bleeding + PAM), power was consistently above 80\% across all scenarios. As expected, the PAM component, being highly discriminative, substantially improved the detectability of treatment effects, confirming the informational gain obtained from richer
composite endpoints.\\

When ICC and CAC were held constant, power increased steadily as $\beta_t$ moved from negative to positive values. Because a positive $\beta_t$ aligns with the direction of the treatment effect in a stepped-wedge design, larger positive trends naturally facilitated detection. On average, power rose from about 30\% when $\beta_t = -0.05$ to around 80\%
when $\beta_t = 0.05$ for the 3 endpoints case, illustrating the strong influence of secular trends on treatment-effect estimation.\\

For fixed $\beta_t$ and ICC, increasing the CAC resulted in a modest but consistent improvement in power. The gain was generally limited to a few percentage points, but it appeared systematically across endpoint sets and methods, reflecting the benefit of stronger autocorrelation structure within clusters.\\

Regarding method performance, the within-cluster probabilistic index model (c2) generally outperformed the mixed-effects comparison model (b4), but the magnitude of this advantage depended strongly on the richness of the composite endpoint. The difference between the
two methods was most noticeable when the composite included two or three components (death~+~VTE, or death~+~VTE~+~bleeding), where c2 achieved clearly higher power. In contrast, when only a single endpoint was used, or when all four outcomes were combined (including PAM), the performance of the two methods was much more similar. Moreover, the advantage of c2 diminished as the temporal trend $\beta_t$ became increasingly favorable
to the treatment: in these scenarios, both methods benefited from the stronger secular signal, reducing the relative gap between them. Overall, c2 remained slightly but consistently superior, though its comparative benefit was concentrated in intermediate endpoint configurations.
\\

Finally, using the operating characteristics obtained from this simulation framework, we assessed the statistical power that would be expected for the ETHER trial under its protocol-defined assumptions. 
In the ETHER design, the intraclass correlation was anticipated to lie between 0 and 0.01, while the cluster autocorrelation was expected to fall between 0.75 and 1. 
Within this range of correlation structures, both candidate methods achieved very high power when applied to the full composite endpoint. 
Across all combinations of $(\text{ICC}, \text{CAC}) \in [0,0.01] \times [0.75,1]$, the estimated power consistently exceeded 90\%, reaching values close to 100\% for the cluster-restricted PIM (c2). 
These results indicate that, under the assumptions of the original ETHER protocol and using the proposed GPC-based analyses, the trial would have more than adequate power to detect the pre-specified treatment effects. 
In particular, the inclusion of the PAM-13 component, which is highly sensitive to treatment changes, plays a decisive role in ensuring high discriminative ability and robust operating performance under realistic trial conditions.

\section{Discussion}

The aim of this work was to investigate whether generalized pairwise comparisons (GPC) can be analysed in a way that appropriately reflects both the correlation structure and the temporal features inherent to stepped-wedge cluster randomised trials (SW-CRTs). As emphasised in the methodological literature, valid inference in SW-CRTs requires statistical approaches that simultaneously adjust for clustering and secular time trends, in order to avoid biased treatment effect estimates \cite{hemming2015}. The results of our simulation study show that such joint adjustment is indeed feasible. In particular, two approaches---Model b4, based on a hierarchical mixed-effects formulation, and Model c2, relying on a cluster-restricted probabilistic index model (PIM)---were identified as consistently providing valid inference while accounting for clustering, repeated measurements across periods, and temporal confounding. These findings align with previous work demonstrating that PIMs can incorporate covariates and structured effects through a regression-based framework \cite{Thas2012}.\\

Among the methods evaluated, Model b4 exhibited excellent robustness across all simulation scenarios, including settings with low cluster autocorrelation (CAC). Its random-effects structure efficiently captures both between-cluster and between-period heterogeneity, making it a particularly reliable option for a wide range of SW-CRT designs \cite{hussey2007}. Model c2 also performed well and often achieved slightly higher statistical power, especially in scenarios with high intracluster correlation (ICC). This gain in efficiency, however, comes at the cost of a substantial computational burden. Because PIMs operate at the level of individual pairwise comparisons, the number of comparisons increases quadratically with the cluster-period sample size, leading to prohibitive runtimes when multiple covariates are included \cite{Thas2012}. In designs with small cluster-period sample sizes, PIMs may therefore remain advantageous, as the total number of pairs is naturally limited. For larger clusters, however, mixed-effects models are clearly more practical.\\

The simulation results further highlight the strong influence of the baseline event probability $p_0$ on both power and type~I error. When $p_0$ was very small (e.g.\ $0.01$), both quantities decreased markedly. This behaviour is expected, as very rare events generate few informative treatment--control comparisons, reducing the ability to detect a true treatment effect while simultaneously lowering the probability of false positives. As $p_0$ increased to more moderate values (such as $0.1$ or $0.2$), the number of informative comparisons increased accordingly, resulting in higher power and type~I error rates closer to their nominal levels.\\

An additional issue concerns small-sample inference for probabilistic index models. PIM estimation relies on the large-sample properties of estimating equations, and when the effective sample size is limited, variance estimators may be biased downward, leading to inflated type~I error rates. This phenomenon has been documented in the methodological literature, and several small-sample corrections for PIMs have been proposed, including bias-corrected sandwich variance estimators and resampling-based approaches \cite{Jaspers}.  

In the present study, this issue was not investigated further, as the simulation scenarios were calibrated to reflect the ETHER trial, which involves a relatively large number of clusters and individuals per period and therefore operates within an asymptotic regime. Nevertheless, in applications with smaller cluster sizes or fewer informative comparisons, the use of such small-sample corrections would be advisable when adopting PIM-based methods.\\

Another important limitation of PIMs is their inability to incorporate random effects, as their estimation relies on marginal estimating equations that do not accommodate hierarchical random-effects structures \cite{Thas2012}. As a consequence, PIMs cannot explicitly model cluster autocorrelation. Attempts to approximate this dependence structure using fixed cluster indicators proved unsuccessful and substantially increased computational cost. In practice, however, the performance difference between Models b4 and c2 remained modest across most realistic scenarios, suggesting that both approaches offer acceptable solutions for the analysis of SW-CRTs.\\

Our simulations primarily focused on linear secular trends, corresponding to a relatively simple temporal structure. When sinusoidal time patterns were explored, the results remained broadly similar, likely because the sinusoidal function oscillates around zero and therefore does not induce strong cumulative trends over time. Future work should consider more complex non-linear temporal structures, such as cubic or spline-based trajectories, which may generate pronounced period-specific biases if not adequately adjusted for.\\

We also explored an extension in which non-linear secular trends are modelled directly within the PIM framework using bivariate spline functions of time. The underlying idea is as follows. Rather than adjusting for time through a linear contrast $(t_{ijk} - t_{i'j'k})$, the observation times $t_{ijk}$ and $t_{i'j'k}$ are first expanded using a spline basis 
$B(t) = \{B_1(t), \dots, B_d(t)\}$ of dimension $d$, such as natural cubic splines. Because PIMs operate on ordered pairs of individuals, temporal adjustment must be expressed not as a univariate function of time, but as a smooth function of the pair $(t_{ijk}, t_{i'j'k})$. This requires constructing a tensor-product spline basis
\[
\Phi(t_{ijk},t_{i'j'k}) = \big\{ B_a(t_{ijk}) \, B_b(t_{i'j'k}) \; : \; 1 \le a,b \le d \big\},
\]
which contains $d^2$ basis functions. Each pair contributes the covariate vector
\[
Z_{ii'} = \big( 1,\; X_{ijk} - X_{i'j'k},\; \Phi(t_{ijk},t_{i'j'k}) \big),
\]
and the PIM estimator solves modified estimating equations of the form
\[
g\left( \mathbb{P}(Y_i \succ Y_i' \mid Z_{ii'}) \right) = Z_{ii'}^\top \beta,
\]
where $g(\cdot)$ typically denotes the logit link function.  

Although this construction offers considerable flexibility for capturing non-linear or irregular temporal dynamics, it is computationally demanding. The number of spline parameters grows quadratically with the spline dimension ($d^2$), and the resulting design matrix rapidly becomes very large. In practice, even moderate spline dimensions (e.g.\ $d=3$) lead to a substantial number of parameters, particularly when combined with the large number of pairwise comparisons inherent to PIM estimation. Consequently, spline-based extensions of PIMs are feasible only for relatively small cluster-period sample sizes or when spline complexity is carefully controlled.\\

In standard stepped-wedge analyses, time is often treated as a categorical factor, reflecting the equal duration of study periods. In the ETHER trial, however, the first period spans 12 months, whereas subsequent periods last only 3 months, creating a clear imbalance. This feature supports modelling time as a continuous variable or through spline-based representations rather than simple categorical indicators. From this perspective, PIMs may offer additional flexibility, as continuous time differences can be incorporated directly at the pairwise level. This advantage could become even more relevant in survival analysis settings, where the timing of events carries particularly rich information.\\

Several aspects remain beyond the scope of this work and merit further investigation. First, unequal cluster sizes---frequent in pragmatic stepped-wedge trials---may interact with cluster-level correlation structures and affect the stability of win-odds estimates. Second, correlations between multiple components of composite endpoints may also influence the variance estimation of GPCs \cite{buyse2010} and deserve dedicated methodological developments. Third, although our simulations confirmed that period-pair comparisons, as implemented in Model b4, perform well, alternative approaches---such as jointly modelling all period contrasts through spline-based adjustments---may further improve efficiency in the presence of complex non-linear time trends.\\

Overall, this study provides practical guidance for the analysis of GPCs in stepped-wedge designs. Mixed-effects modelling (Model b4) emerges as a generally reliable and scalable approach, while PIM-based methods (Model c2) can offer slightly higher efficiency under certain conditions, at the expense of substantial computational cost and without the ability to incorporate random effects. In most realistic scenarios, both approaches lead to comparable conclusions, and the choice between them should therefore be guided by considerations of computational feasibility, cluster-period sample size, and the anticipated complexity of temporal trends.

\bmsection*{Funding/Support}

The study is funded by the European Union grant HORIZON-
HLTH-2022-TOOL-11-01. Views and opinions expressed are however those of the author(s)
only and do not necessarily reflect those of the European Union or HORIZON-HLTH-2022-
TOOL-11-01. Neither the European Union nor the granting authority can be held responsible
for them.

\bmsection*{Conflict of interest}

The authors declare no potential conflict of interests.

\bibliography{sample}

\bmsection*{Supporting information}

Additional supporting information may be found in the
online version of the article at the publisherâs website.

\appendix


\end{document}